\documentclass[twoside]{article}
\usepackage{fleqn,espcrc2}
\usepackage{epsf}
\usepackage{epsfig}

\usepackage{graphicx}

\usepackage[figuresright]{rotating}

\newcommand{\fnm}[1]{\footnotemark[#1]}

\title{
Interpretation of a microwave induced current step in a single intrinsic 
Josephson junction on a Bi-2223 thin film}

\author{Ch. Helm\address{Los Alamos National Laboratory, T-11, M.S. B-262, NM 
87545, USA and \newline
Institute of Theoretical Physics, University of Regensburg, D-93040 Regensburg,
Germany}, 
A. Odagawa\fnm{2}, M. Sakai\fnm{2}, H. Adachi\fnm{2}, 
K. Setsune\address{Advanced Technology Research Laboratory, Matsushita 
Electric Industrial Co., Ltd., Seika, Soraku, Kyoto, 619-0237 Japan} and
R. Kleiner\address{Physikalisches Institut, Lehrstuhl Experimentalphysik II, 
Universit{\"a}t T{\"u}bingen, 
aAuf der Morgenstelle 14, 
D-72076 T{\"u}bingen, 
Germany}}

\begin{document}

\begin{abstract}
Thin stacks consisting of a single intrinsic Josephson
junction on (Bi,Pb)-Sr-Ca-Cu-O thin films are investigated
under the influence of external microwave fields. 
The $I$-$V$-characteristic shows a single resistive branch, a clear 
superconducting gap edge
 structure and a pronounced current step in external microwave fields. 
With increasing irradiation power it shifts to higher voltages, 
while the height of the step remains practically unchanged. 
In a numerical simulation 
including an ac-magnetic field parallel to the superconducting layers
the experimental features of the structure can be explained by a collective
motion of Josephson fluxons. 
\end{abstract}

\maketitle




\section{INTRODUCTION}

It is well established now that the electronic $c$-axis transport 
in the superconducting state of high-$T_c$-superconductors (HTSC) like 
$(Bi,Pb)_2Sr_2Ca_2Cu_3O_{10+x}$ (Bi-2223) is determined by an intrinsic 
Josephson effect between the superconducting $CuO_2$-layers  
\cite{kleiner1,tachiki,matsuda,kleiner2,wir1}. Recently, the microwave
properties 
of these materials attracted considerable attention for future electronical 
applications like high frequency oscillators
and high-speed digital devices. 
In this context microwave phase-locking \cite{irie}, the emission of Cherenkov
radiation \cite{hechtfischer} and collective fluxon motion \cite{krasnov}
have been reported. However, due to the strong (inductive) coupling of the 
dynamics in different intrinsic Josephson junctions (ITJJ) 
a detailed analysis is complex. 
Therefore the study of recently fabricated samples consisting of a 
{\em single} junction is interesting, as it rules out the 
influence of interactions and heating effects.

\section{EXPERIMENTAL RESULTS}

We have prepared Bi-2223 thin films using Pb dopant as a stabilizer 
of the 2223 phase on a MgO(100) substrate. Details of the preparation 
technique and the experimental setup can be found elsewhere 
\cite{adachi,odagawa2}. 
SEM, AFM and TEM analysis reveal that the prepared films were composed of 
crystal grains with a size of  $4 \mu {\rm m} \times 4 \mu {\rm m}$ and a 
roughness
of about $1.8 {\rm nm}$, which is one half of the $c$-axis lattice constant. 
The mesa-type stack structures with an area of $\sim 2 \mu {\rm m} \times 2 
\mu {\rm m}$ were fabricated using standard photolithography and $Ar$-ion 
milling techniques.
The $I$-$V$-curve of the stacks show 1-3 branches, which is 
consistent with the height of the stack as expected from the etching 
rate of $\sim 10 {\rm nm}/{\rm min}$. 

Typical parameters of the junctions are as
follows: critical temperature $T_c \sim 100 K$, critical current density $j_c 
\sim 4 \cdot 10^3 {\rm A}/{\rm cm}^2$. In contrast to stacks with a lot 
of ITJJs, it is possible to extract the normal junction resistance 
$R_n \sim 32 
\Omega$, the $I_c R_n\sim 5.1 {\rm mV}$  product and the value
$\Delta_0 \sim 37.5 {\rm mV}$ of the superconducting gap directly. Due to the 
nonlinearity 
of the $I$-$V$-curve the characteristic frequency $f_c$ and the McCumber 
parameter $\beta_c$ are 
strictly speaking not well defined material parameters, but 
they can be roughly estimated from $R_n$ 
($f_c = 2eI_cR_n/h \sim 2.5 {\rm THz}$)
or from the return current $I_r$ (from the resistive to the superconducting 
state): 
$\beta_c = (4 I_c / \pi I_r )^2 \sim 16.6$ \cite{likharev}. 
The contact resistance of the Au/Bi-2223 interface
was evaluated by the slope  of the superconducting branch at $4.2 {\rm K}$
 as $\sim 8 \cdot 10^{-8} \Omega {\rm cm}^2$. 
This value for the contact resistance is several orders of magnitude smaller 
than in previous experiments \cite{kleiner1,kleiner2} and essentially 
eliminates possible nonequilibrium effects due to heating.

\begin{figure}
\caption{\label{ivcurve_general} 
$I$-$V$-characteristic at $4.2 {\rm K}$ of a mesa with dimensions of
$2 \mu {\rm m} \times 2 \mu {\rm m}$ showing a single ITJJ. The experimental 
curve coincides with the theoretical one assuming a $d$-wave order parameter.
}
\end{figure}

\begin{figure}[th!]
\begin{center}
\end{center}
\vspace{-0.9cm}
\caption{\label{IVstep_power} 
(top) Dependence of the current step structure on the 
microwave power $P$, which gradually increases from (a) to (d) 
($5 {\rm mV}/{\rm DIV}$ for voltage and $50 \mu {\rm A}/{\rm DIV}$ for 
currents). \newline 
(below) Voltage position $V_m$ (a)  and height $I_m$ (b)
of the current step as a function of $\sqrt{P}$. 
}
\vspace{-0.5cm}
\end{figure}

Figure \ref{ivcurve_general} shows the $I$-$V$-characteristic at 4.2 K of the
thinnest fabricated stack with a single resistive branch
corresponding to a single unit cell.  The
 $I$-$V$-curve can be well reproduced theoretically assuming 
a $d$-wave order parameter $\Delta (\theta) = \Delta_0 \sin (2 \theta) $
and a parallel resistance of $7.8 \Omega$.

%
%

%
%
%
%
%

We have also studied the behaviour of the single ITJJ 
in external microwave  radiation in the 
frequency range up to $f_c \sim 27 {\rm GHz}$, which is 
two orders of magnitude smaller than 
the characteristic frequency $f_c \sim 2.5 {\rm THz}$. 
In this case the maximum current $I_{s, {\rm max}}$
of the superconducting branch decreased monotonically with increasing 
microwave power \cite{krasnov}. 

In addition to this, we observed a pronounced current step structure 
in the $I$-$V$-curve at a certain voltage $V_m$. 
For amplitudes, which suppress the 
superconducting branch completely, this structure appears in the lower 
millivolt range. 
Figure \ref{IVstep_power} 
shows that with increasing power $P$ of the microwave irradiation 
the step structures shift to higher voltages 
$V_m \sim \sqrt{P}$, while the height 
$I_m$ of the step remains practically unchanged. 


\section{DISCUSSION}

Shapiro steps can be ruled out as an explanation of this phenomenon, because 
they are expected at  voltages $V_{{\rm sh},n} = n \hbar \omega_{\rm rf}/2e$,
 which 
are integer multiples of the voltage $\omega_{\rm rf}/2e \sim 34 \mu V$. 
As the difference of their  voltages 
is two orders of magnitude smaller than $V_m$, this 
interpretation becomes very unlikely, as it would predict  a series of 
high-order Shapiro steps with rather low amplitude instead of one pronounced 
step. 

On the other hand, the lateral size $L_{ab} \sim 2 \mu m$ of the stack is
still larger than the typical size $2 \lambda_J \le 0.6 \mu {\rm m}$ 
\cite{kleiner_vortex} of a Josephson vortex, which allows for fluxon motion
parallel to the layers.
It is thus reasonable to associate the observed step structure with some kind 
of collective vortex flow.

\begin{figure}[ht]
\raisebox{-0.13\textheight}
{
}
\unitlength1cm
\begin{minipage}{0.24\textwidth}
\end{minipage}
%
%
%
%
%
%
%
%
%
%
\caption{\label{IV_comparison} 
Comparison of  experimental (left) and theoretical (right) $I$-$V$-curve of 
the single ITJJ with $f_{\rm rc} \approx 27 {\rm GHz}$ 
for different amplitudes $H_{ac}$ of the ac-magnetic field
($\gamma= I/I_c$, $\nu = \hbar / 2e \langle {\dot \Phi}  \rangle_t$, 
$H_{ac0}=3.5/5 {\hat =} 2/2.8T$). 
}
\end{figure}





%
%

%
%
%

%
%

A complete model of the phase dynamics under the influence of external 
microwave irradiation would include the detailed discussion of the 
(unknown) boundary conditions on the surface of sample in order to 
determine the magnitude of the induced electric and magnetic fields both 
parallel and perpendicular to the superconducting layers. 
Also various pinning mechanisms should in principle be taken into account. 

For the reproduction of the experimental features presented above it will 
be sufficient to consider the effect of ac-magnetic fields 
$H_{\rm ac} (t) = H_{\rm ac0} \sin (\omega_{\rm rf} t)$. 
Note that the influence of an external magnetic field $H_{\rm ac}$ is 
formally equivalent to currents injected parallel to the layers. 
It also turns out in the simulation that an externally applied oscillating 
$c$-axis current $I_{\rm ac} (t)$ is unable to reproduce the experimental 
data correctly. 
The nonlinearity of the quasiparticle current is modelled as in 
\cite{wir1,schlenga}.

\begin{figure}
%
%
%
\caption{\label{snapshot_A} 
Snapshots of the simulated supercurrent distribution $\sim \sin \Phi (x,t)$
in region A/B (left/right) (cf. Fig. \ref{IV_comparison}). 
Open/closed circles represent 
the center of fluxons/anti-fluxons moving to the right/left 
respectively. The plots (a)-(e) 
are taken at different (not equidistant) time steps ($H_{\rm ac0} =2 T$). 
}
\end{figure}

Figure \ref{IV_comparison}  compares the 
experimental step structure and the theoretical simulation using the parameters
given above. Thereby both the dependence of the critical current $I_c$ and 
of the voltage $V_m$ and the height $I_m$ on the external microwave power 
$P \sim H_{ac0}^2$ can be successfully reproduced. 

As the typical frequencies used here are much smaller than the plasma 
frequency, the external microwaves have similar effects as static external 
fields and consequently do not depend on the exact value of the oscillation 
frequency. 

As a consequence, the behaviour of the observed structure can be understood 
in terms of well known features of the flux-flow step in high magnetic 
fields \cite{hechtfischer,krasnov}: $V_m = B s L_{ab}$ ($s$: distance of 
superconducting layers) and $I_m/I_c= c_s/L_{ab} f_c \sim 0.06$. 
Physically the step structure occurs when the Josephson vortices created by 
the external field approach their limiting Swihart velocity ${\bar c} \approx 
2-5 \times 10^5 {\rm m}/{\rm s}$ \cite{hechtfischer,sakai}. 


The numerical solution of the Sine-Gordon equation allows to discuss the 
supercurrent distribution $I_c \sin \Phi$  in the regions A and B as marked 
in Fig. \ref{IV_comparison}. 
Open/closed circles in Fig. \ref{snapshot_A} represent the center of 
vortices/anti-vortices respectively and the plot (a)-(e) are taken at 
different (not equidistant) time steps. In contrast to the static case, 
the direction of moving fluxons will change with the external frequency
$\omega_{rf}$, changing the polarity according to the (alternating) 
direction of the external field $H_{ac} (t)$. 

In region A (cf. Fig. \ref{snapshot_A})  the phase is increasing linearly 
$\Phi(x,t) \sim \Phi_0 + k x$, which corresponds to a very homogeneous 
field distribution in the stack. 
On the other hand, in region B more pronounced 
kinks in the phase can be found, which account for a loose array of 
vortices with periodically oscillating relative distance of the fluxons.

\section{CONCLUSIONS} 

The successful fabrication of single intrinsic junction stacks on 
Bi-2223 thin films  with  low contact resistance and 
$I_c R_n \sim 5.1 {\rm mV}$ has been reported. 
Due to this fact we were able to study the properties of a single junction 
without interference with different junctions in the stack and eliminating 
the influence of heating. 
Under microwave irradiation, we observed a pronounced
step in the $I$-$V$-characteristic of the single ITJJ in the 
lower millivolt range well below the superconducting gap edge. Its voltage 
position changes linearly as a function of the square root of the irradiated 
microwave power, while the step current remains  constant.
This behaviour could be qualitatively reproduced by numerical simulations 
in an external ac-magnetic field parallel to the layers, which show 
a collective motion of vortices in alternating directions. 



The authors would like to thank Dr. K. Mizuno for valuable discussions.
One of us (C.H.) gratefully acknowledges the hospitality of the Advanced 
Technology Research Laboratories in Kyoto 
and financial support by JISTEC, the Studienstiftung
des Deutschen Volkes and the Department of Energy under contract 
W-7405-ENG-36.


\begin{thebibliography}{9}

\bibitem{kleiner1}
R. Kleiner, F. Steinmeyer, G. Kunkel, P. M{\"u}ller, Phys. Rev. Lett.
 {\bf 68} (1992) 2394.
\bibitem{tachiki} M. Tachiki, T. Koyama, S. Takahashi, Phys. Rev. B {\bf 50}
 (1994) 7065.
\bibitem{matsuda} Y. Matsuda, M.B. Gaifullin, K. Kumagai, K. Kadowaki, 
T. Mochiku, Phys. Rev. Lett. {\bf 75} (1995) 4512.
\bibitem{kleiner2}
R. Kleiner, P. M{\"u}ller, Phys. Rev. B 49 (1994) 1327.
\bibitem{wir1}
Ch.~Helm, Ch. Preis, F. Forsthofer, J. Keller,
K. Schlenga, R. Kleiner,  P. M\"ul\-ler,
Phys. Rev. Lett. {\bf 79} (1997) 737.
\bibitem{irie}
A. Irie, G. Oya, IEEE Trans. Appl. Supercond. 5 (1995) 3267.
\bibitem{hechtfischer}
G. Hechtfischer, R. Kleiner, K. Schlenga, 
W. Walkenhorst, P. M{\"u}ller, Phys. Rev. B 55 (1997) 14638.
\bibitem{krasnov} V.M. Krasnov, N. Mros, A. Yurgens, D. Winkler, 
Phys. Rev. B {\bf 50}, 13 (1999) 8463. 
\bibitem{adachi} 
H. Adachi, Y. Ichikawa, K. Setsune, S. Hatta, K. Hirochi, K. Wasa, Jpn. 
J. Appl. Phys. {\bf 27} (1988) L643.
\bibitem{odagawa2}
A. Odagawa, M. Sakai, H. Adachi, K. Setsune, Jpn. J. Appl. Phys. {\bf 37}
(1998) 486.
\bibitem{likharev}
K.K. Likharev, Dynamics of Josephson junctions and circuits, 
Gordon and Breach, Philadelphia, 1986.
\bibitem{kleiner_vortex} R. Kleiner, P. M{\"u}ller, H. Kohlstedt, N.F. 
Pedersen, S. Sakai, Phys. Rev. B {\bf 50} (1995), 3942.
\bibitem{schlenga} K. Schlenga, R. Kleiner, G. Hechtfischer, 
 M. M{\"o}{\ss}le, S. Schmitt, P. M{\"u}ller, Ch. Helm, Ch. Preis, 
F. Forsthofer, J. Keller, H.L. Johnson, M. Veith, E. Steinbei{\ss}, Phys. 
Rev. B {\bf 57} (1998) 14518. 
\bibitem{sakai}
S. Sakai, P. Bodin, N.F. Pedersen, J. Appl. Phys. {\bf 73} (1993) 2411.


\end{thebibliography}
\end{document}